\documentclass[fleqn,usenatbib]{mnras}

\pdfoutput=1  

\usepackage{graphicx}            
\usepackage{amsmath}             
\usepackage{amssymb}             
\usepackage{gensymb}             
\usepackage{bm}                  
\usepackage{xspace}              
\usepackage{nicefrac}            
\usepackage{hyperref}            
\usepackage{xcolor}              
\usepackage{ifthen}
\usepackage{enumitem}
\usepackage{subfig}
\usepackage{suffix}
\usepackage{textcomp}
\usepackage{fontawesome}

\title[Matching observed and simulated lensing galaxies]{A new
  strategy for matching observed and simulated lensing galaxies}

\newcommand{\nauthor}[2]{#2,$^{#1}$}
\newcommand{\nauthorl}[2]{#2$^{#1}$}
\newcommand{\email}[1]{\thanks{Email: #1}}
\newcommand{\naffiliation}[2]{$^{#1}$#2}

\newcommand\SDSSJ[1]{\ifthenelse{\equal{#1}{0029}}{\textsc{J0029\textminus0055}}{}\ifthenelse{\equal{#1}{0737}}{\textsc{J0737+3216}}{}\ifthenelse{\equal{#1}{0753}}{\textsc{J0753+3416}}{}\ifthenelse{\equal{#1}{1051}}{\textsc{J1051+4439}}{}\ifthenelse{\equal{#1}{0956}}{\textsc{J0956+5100}}{}\ifthenelse{\equal{#1}{1430}}{\textsc{J1430+6104}}{}\ifthenelse{\equal{#1}{1627}}{\textsc{J1627\textminus0053}}{}\ignorespaces}%

\newcommand{\kms}{\ensuremath{\mathrm{kms^{-1}}}\xspace}

\newcommand\roche{\mathcal{P}}

\newcommand{\seclbl}[1]{\label{sec:#1}}
\newcommand{\subseclbl}[1]{\label{subsec:#1}}
\newcommand{\figlbl}[1]{\label{fig:#1}}
\newcommand{\tablbl}[1]{\label{tab:#1}}
\newcommand{\eqlbl}[1]{\label{eq:#1}}

\newcommand{\secref}[1]{Section~\ref{sec:#1}}
\WithSuffix\newcommand\secref*[1]{Section~\ref{sec:#1}}
\newcommand{\subsecref}[1]{Subsection~\ref{subsec:#1}}
\WithSuffix\newcommand\subsecref*[1]{\ref{subsec:#1}}
\newcommand{\figref}[1]{Figure~\ref{fig:#1}}
\WithSuffix\newcommand\figref*[1]{\ref{fig:#1}}
\newcommand{\tabref}[1]{Table~\ref{tab:#1}}
\WithSuffix\newcommand\tabref*[1]{\ref{tab:#1}}
\renewcommand{\eqref}[1]{Eq.~(\ref{eq:#1})}
\WithSuffix\newcommand\eqref*[1]{(\ref{eq:#1})}

\newcommand{\Code}[1]{\texttt{#1}}

\newcommand*{\home}{.}%

\author[Denzel et al.]{%
  \nauthor{1,2}{Philipp Denzel} \email{phdenzel@physik.uzh.ch}
  \nauthor{3}{Sampath Mukherjee} 
  \nauthorl{1,2}{Prasenjit Saha} 
  \newauthor
  \\
  \naffiliation{1}{Institute for Computational Science, University of Zurich, CH-8057 Zurich, Switzerland} \\
  \naffiliation{2}{Physics Institute, University of Zurich, CH-8057 Zurich, Switzerland} \\
  \naffiliation{3}{STAR Institute, Quartier Agora - All\'ee du six Ao$\hat{u}$t, 19c B-4000 Li\`ege, Belgium} \\
}

\date{}
\pubyear{}

\begin{document}

\label{firstpage}
\pagerange{\pageref{firstpage}--\pageref{lastpage}}

\maketitle

\begin{abstract}
\noindent The study of strong-lensing systems conventionally involves
constructing a mass distribution that can reproduce the observed
multiply-imaging properties.  Such mass reconstructions are
generically non-unique.  Here, we present an alternative strategy:
instead of modelling the mass distribution, we search cosmological
galaxy-formation simulations for plausible matches.  In this paper we
test the idea on seven well-studied lenses from the SLACS survey.  For
each of these, we first pre-select a few hundred galaxies from the
EAGLE simulations, using the expected Einstein radius as an initial
criterion.  Then, for each of these pre-selected galaxies, we fit for
the source light distribution, while using MCMC for the placement and
orientation of the lensing galaxy, so as to reproduce the multiple
images and arcs.  The results indicate that the strategy is feasible,
and even yields relative posterior probabilities of two different
galaxy-formation scenarios, though these are not statistically
significant yet.  Extensions to other observables, such as kinematics
and colours of the stellar population in the lensing galaxy, is
straightforward in principle, though we have not attempted it yet.
Scaling to arbitrarily large numbers of lenses also appears feasible.
This will be especially relevant for upcoming wide-field surveys,
through which the number of galaxy lenses will rise possibly a
hundredfold, which will overwhelm conventional modelling methods.
\end{abstract}


\section{Introduction}\seclbl{match:intro}

Four decades after the first discovery by \cite{Walsh1979}, galaxies exhibiting
strong gravitational lensing seem almost commonplace.  The SLACS sample
\citep[Sloan Lens ACS;][]{SLACS1, SLACS5, SLACS13} alone has over a hundred
strong lensing galaxies.  The next generation of wide-field surveys (LSST/Rubin
from the ground, and Euclid and WFIRST/RST in space) promise many many more.
Extrapolation from small fields that have been surveyed at different resolutions
indicate \citep[see e.g.,][]{Collett15} that $>100'000$ strong-lensing galaxies
await discovery.  Many techniques for finding lenses in surveys, ranging from
crowdsourcing \citep{SW1} to neural networks \citep[e.g.,][]{Davies19}, have
been developed in recent years, and one can confidently expect that $10^5$
strong-lensing galaxies will be discovered.

Meanwhile, the past decade has seen significant progress on the structure and
formation of galaxies.  Within the $\Lambda$CDM paradigm, there is general
agreement regarding the growth of density perturbations under gravity, from the
level observed in the cosmic microwave background to the formation of
dark-matter halos.  The subsequent processes of star formation and the resultant
feedback are less well understood and require sub-grid models to simulate, but
still the galaxies formed in simulations like Illustris \citep{Illustris}, FIRE
\citep[Feedback In Realistic Environments;][]{FIRE}, and EAGLE \citep[Evolution
and Assembly of GaLaxies and their Environments;][]{Crain15} are much more
credible than previous generations of simulated galaxies.  The \Code{SEAGLE}
pipeline \citep[Simulating EAGLE LEnses;][]{SEAGLE-I} producing simulated lenses
from EAGLE is of particular interest in this work.  In addition to
galaxy-formation simulations, there are also distribution-function models for
galaxies, such as from \Code{AGAMA} \citep[action-based galaxy modelling
architecture; ][]{AGAMA}, which provide self-consistent phase-space
distributions for dark matter, stars and gas.

One would like to compare lensing observations with galaxy simulations.  Let us
first consider this task in a rather abstract way.  Let $F$ be some
galaxy-formation scenario, and let $D$ represent the observational data.  In
Bayesian terms, the posterior probability of $F$ after comparison with $D$ would
be
\begin{equation}
  P(F\,|\,D) = \frac{P(D \,|\, F) \, P(F)} {P(D)}
\end{equation}
where $P(F)$ represents the probability of $F$ before the data, and $P(D)$ is
the probability of the data marginalized over all possible $F$.  The factors
$P(F)$ and $P(D)$ cancel if we compare two formation scenarios with equal prior
probability, so it is really $P(D\,|\,F)$ that is of interest.  This quantity is
given by the marginalisation
\begin{equation}\eqlbl{match:bayes_concept1}
  P(D\,|\,F) = {\textstyle\sum_g}\, P(D \,|\, g) \, P(g \,|\, F)
\end{equation}
where $g$ represents galaxy properties.  There will also be nuisance parameters
(call these $\nu$), such as the orientation of the ellipticity of a galaxy,
which are also to be marginalised over, thus
\begin{equation}\eqlbl{match:bayes_concept2}
P(D\,|\,g) = {\textstyle\sum_\nu}\, P(D \,|\, g,\nu) \, P(\nu) \,.
\end{equation}

Conventional lens modelling consists of constructing $g$ so as to
optimise $P(D\,|\,g)$.  Here, there are two basic approaches.  One is
to assume some parametric form for the lensing mass distribution and
fit to the data.  The idea goes back to the very first lens-modelling
paper \citep{Young1980}.  Recent parametric lens models \citep[such
  as][]{GLEE} are more elaborate, but still much simplified compared
to a simulation from \Code{AGAMA} or \Code{SEAGLE}.  Alternatively,
one can let the lensing mass distribution be free-form, and sample the
abstract space of mass distributions that fit the data.  This approach
is more common in cluster lensing \citep[see e.g.,][]{Wagner19}, but
also used in galaxy lensing \citep[e.g.,][]{Kueng18}.  Free-form mass
models are more complex, but they are not necessarily dynamically
plausible.  Neither style of lens modelling has much input from
$P(g\,|\,F)$.  Some comparisons of lens models against dynamical
simulations of galaxies have been done \citep[e.g.][]{Saha06c,
  CAULDRONtest, GLASS, TDLMC2}, as have some model-independent
comparisons of image statistics with substructure in $\Lambda$CDM
\citep{Gomer17}, but all of these provide only qualitative information
with respect to $P(g\,|\,F)$.

In this work, we attempt for the first time a direct comparison of lensing data
and galaxy-formation without conventional lens models.  We use \Code{SEAGLE}
lenses as samples of $P(g\,|\,F)$ from two different galaxy-formation scenarios.
We then formulate $P(D\,|\,g,\nu)$ so that a procedure for fitting source
brightness distributions \citep[developed earlier for conventional lens
modelling][]{Denzel20,Denzel20b} can be repurposed.  This allows us to find
EAGLE galaxies that can account for the observed images in a small test sample
of seven SLACS lenses (see \tabref{match:systems}).  As this work is intended as
proof of concept, we do not include data other than multiple images from
extended sources.

The following \secref{match:methods} introduces what we may call the
$P(D\,|\,F)$ method.  The subsequent \secref{match:seagle} details the
\Code{SEAGLE} pipeline and summarizes how the catalogue of surface-density maps
was compiled.  The selected test-case lenses from the SLACS survey are presented
in \secref{match:testcases}, and the results of these tests are reported in
\secref{match:results}.  Finally, a summary and discussion, in particular about
possible expansions and applications of the lens-matching approach are given in
\secref{match:conclusion}.

\section{The plausible-match method}\seclbl{match:methods}

To go beyond the simple abstractions above and discuss the actual method, let us
rewrite Eqs.~\eqref*{match:bayes_concept1} and \eqref*{match:bayes_concept2} as
\begin{equation}\eqlbl{match:bayes_concept}
  P(D\,|\,F) \approx
  \sum_{s,\xi}
  P\big(I^{\rm obs}\,\big|\,\bm\alpha,s\big) \,
  P\big(\bm\alpha,\big|\,\xi,F\big) \,
  P(s,\xi) \,.
\end{equation}
Rather than galaxy properties $g$ in general, we are concerned with a lensing
deflection field $\bm\alpha$.  The $\nu$ parameters consist of (a)~location and
rotation parameters (say $\xi$) to produce $\bm\alpha$ from a simulated EAGLE
galaxy, and (b)~the unlensed brightness distribution $s$ at the source redshift.
The priors $P(s,\xi)$ we take as flat.  Hence it is on the factors
$P(D\,|\,\bm\alpha,s)$ and $P(\bm\alpha\,|\,\xi,F)$ that we must concentrate.

\subsection{Data adaptation}\subseclbl{match:data_scaling}

We now describe the ingredients for the factor
$$ P\big(\bm\alpha,\big|\,\xi,F\big) $$
in \eqref{match:bayes_concept}.

The convergence map (that is, the lensing mass distribution in dimensionless
form) is given by the usual projection of the 3D mass density as
\begin{equation}\eqlbl{match:thinlens}
  \kappa(\bm\theta,\xi) = \frac{4\pi G}{cH_0}\,
  \frac{d_\mathrm{LS}d_\mathrm{L}}{d_\mathrm{S}}
  \int \rho(\bm\theta,\xi,z)\,\mathrm{d}z \,.
\end{equation}
Here, $\bm\theta$ is the angle on the observer's sky, $d_{LS}$ is the
dimensionless angular-diameter distance from the lens to the source, $d_L$ and
$d_S$ are analogous, and $\xi$ represents the location and orientation of the 3D
density $\rho(\bm\theta,z)$.  A conventional $\Lambda$CDM cosmology is assumed.

In this work, we have limited the analysis to two galaxy-formation scenarios
from the EAGLE simulations (details are in \secref{match:seagle} below).  From
the two simulations, \Code{SEAGLE} projected each of 554 simulated galaxies
along three orthogonal axes to produce maps of $\kappa(\bm\theta)$ for the
fiducial redshift values $z_L = 0.23$ and $z_S = 0.8$.  The $\kappa$ maps have
$161\times161$ square pixels with a pixel size of $0.05\arcsec$, yielding an
angular size of about $8\arcsec\times8\arcsec$ for an entire map.  The $\kappa$
distributions differ in size and in shape.  \figref{match:catalogue} shows the
distribution of mean enclosed $\kappa$ as a function of radial distance from
lens centre for the entire catalogue.  The notional Einstein radius is the value
of $\theta_E$ for which $\langle\kappa\rangle_{\theta_E} = 1$.  About 20 mass
maps in the catalogue are always below $\kappa=1$ and hence are not strongly
lensing for $z_L = 0.23, z_S = 0.8$.  But most of the galaxies can produce
multiple images, and the Einstein radii go up to $3.15\arcsec$.

\begin{figure}%
    \centering%
    \includegraphics[width=0.475\textwidth]{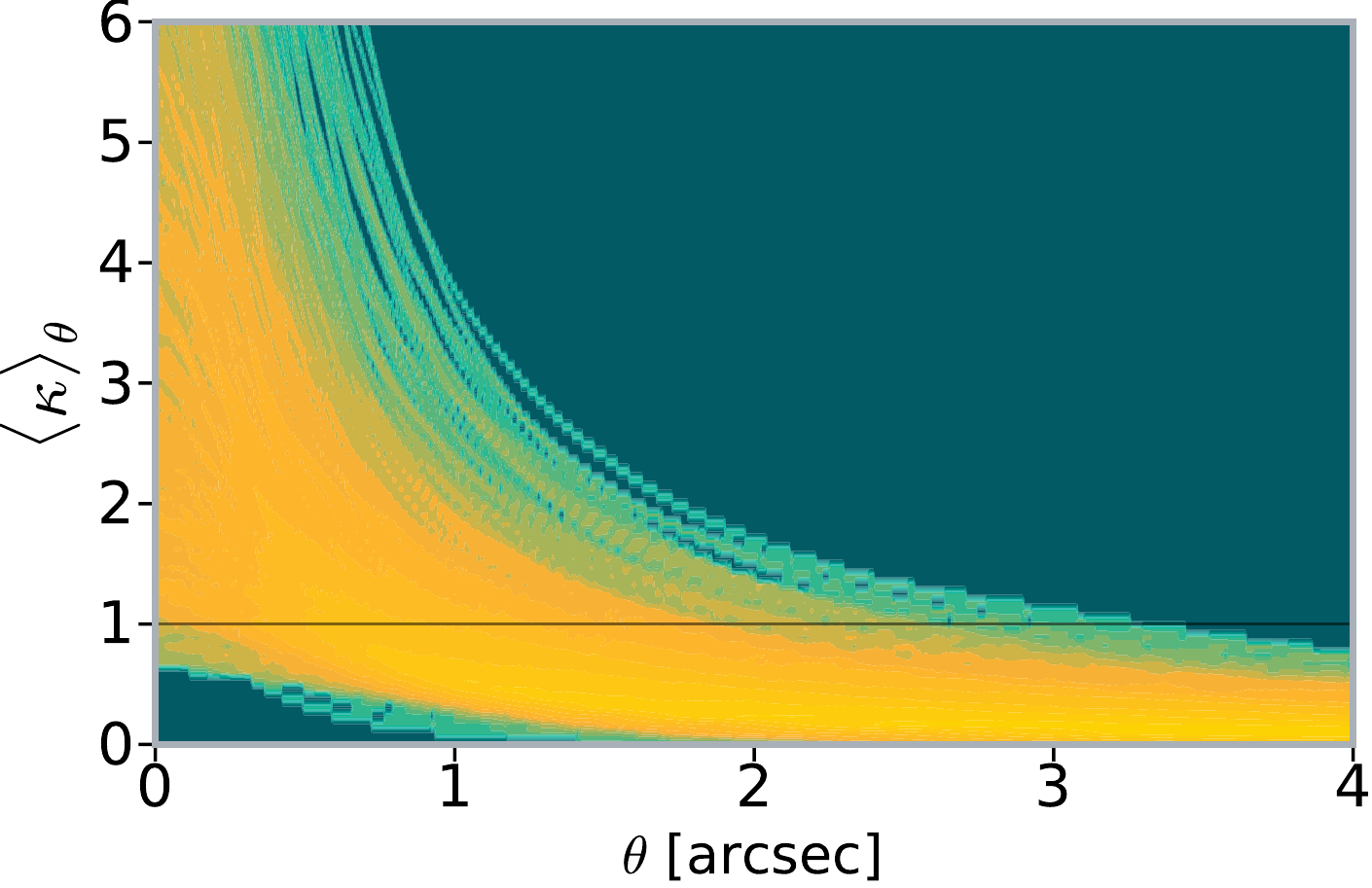}%
    \caption{The distribution of radial profiles of the mean enclosed
      $\langle\kappa\rangle$ within a given projected radius from the
      centre of the galaxies, for the entire catalogue, assuming $z_L
      = 0.23,z_S = 0.8$.  The horizontal line at $\kappa = 1$
      indicates the notional Einstein radii.}\figlbl{match:catalogue}%
\end{figure}%

The convergence maps are then rescaled from $z_L=0.23,z_S=0.8$ to the redshift
values corresponding to each of the test-case lens system listed in
\tabref{match:systems}.  A subset of a few hundred $\kappa$ maps with Einstein
radius in the expected range for each system was then selected for further
processing.

The next step was the computation of the lens potential $\psi(\bm\theta) =
2\nabla^{-2}Q(\bm\theta)$.  To reduce the computational time required, the
$\kappa$ maps were discretised to $23\times23$ tiles.  The potential is then
expressed as
\begin{equation}\eqlbl{match:pixelized_potential}
\psi(\bm\theta) = 2\sum_{n}\kappa_{n}\nabla^{-2}Q(\bm\theta - \bm\theta_n)
\end{equation}
where $\kappa_n$ is the density of the $n$-th tile and $Q(\bm\theta -
\bm\theta_n)$ is the contribution of a square tile with constant $\kappa=1$
located at $\bm\theta_n$.  The functional form of $Q(\bm\theta)$ is given in
\cite{Abdelsalam98}.  Note that only the mass distribution is reduced in
resolution in this way, but $\bm\theta$ and $\psi(\bm\theta)$ can still be
evaluated at any desired resolution.  The effect of the approximation
(\ref{eq:match:pixelized_potential}) is expected to be very small.

Once the lens potential is known we have the deflection angle as
\begin{equation}\eqlbl{match:deflection}
  \bm\alpha(\bm\theta) = \nabla\psi(\bm\theta) \,.
\end{equation}

\subsection{Synthetic images}\subseclbl{match:synthetics}

We now consider the factor
$P\big( I^{\rm obs} \,\big|\, \bm\alpha,s \big)$
in \eqref{match:bayes_concept}.

As a result of the deflection (\ref{eq:match:deflection}) a light ray
originating at a source at $\bm\beta$ on the sky will be observed at $\bm\theta$
which is related to $\bm\beta$ by the usual lens equation
\begin{equation}\eqlbl{match:lens_equation}
    \bm\beta = \bm\theta - \bm\alpha(\bm\theta) \,.
\end{equation}
The lens equation amounts to a mapping $L(\bm\theta,\bm\beta)$ between the
source and image planes, which can be discretised as a matrix.  Any given
$\bm\theta$ corresponds to a unique $\bm\beta$, whereas a given $\bm\beta$ may
correspond to more than one $\bm\theta$.  A source-brightness distribution
$s(\bm\beta)$ produces an image-brightness distribution
\begin{equation}\eqlbl{match:mapping}
I(\bm\theta) = \int L(\bm\theta, \bm\beta) \, s(\bm\beta) \,
\mathrm{d}^2\bm\beta \,.
\end{equation}
The observed image brightness will involve a further convolution with the
point-spread function (PSF) $P(\bm\theta - \bm\theta')$ of the telescope and
camera.  The result
\begin{equation}\eqlbl{match:synthetics}
\bar I(\bm\theta) = \int P(\bm\theta - \bm\theta') \, I(\bm\theta') \,
\mathrm{d}^2\bm\theta'
\end{equation}
we will call the synthetic image, and it is what will get compared with the
data.

For the lens sample investigated here, appropriate PSFs have been employed which
were modelled using \Code{tinytim}\footnote{\faGithub\,
\url{https://github.com/spacetelescope/tinytim}} \citep{tinytim}.

Assuming now that the detector noise is Gaussian with known $\sigma_{\bm\theta}
\propto \sqrt{I^{\rm obs}(\bm\theta)}$ we take
\begin{equation}
P\big( I^{\rm obs} \,\big|\, \bm\alpha,s \big)
\propto \exp\left(-{\textstyle\frac12}\chi^2\right)
\end{equation}
where
\begin{equation}\eqlbl{match:synthetics_chisqr}
\chi^2 = \sum_{\bm\theta} \sigma_{\bm\theta}^{-2}
  \left[I^{\rm obs}(\bm\theta) - \bar I(\bm\theta,\bm\alpha,s)\right]^2
\end{equation}

From Eqs.~(\ref{eq:match:mapping}) and (\ref{eq:match:synthetics}) it is clear
that the synthetic image $\bar I(\bm\theta)$ is linear in the source-brightness
distribution, even though it is completely non-linear in the mass distribution.
Hence $s(\bm\beta)$ can be solved to optimise $P\big( I^{\rm obs} \,\big|\,
\bm\alpha,s \big)$ by linear least-squares.  It is important, however, to mask
the light from the lensing galaxy, since it is not part of
$I^{\text{obs}}(\bm\theta)$.

As the lensed images are typically highly magnified, the source or
$\bm\beta$ plane needs much smaller pixels than the image or
$\bm\theta$ plane.  For this reason, the lens mapping
$L(\bm\theta,\bm\beta)$ maps each $\bm\theta$ pixel to a cluster of
$\bm\beta$ pixels.  To simplify the computation, we replace each
$\bm\theta$ pixel by its central point for the purposes of the lens
mapping.  Then each image pixel maps to a single source pixel.  This
procedure leaves many source pixels `blank', because they send light
to edges and corners of the image pixels.  These blank pixels could be
filled in by interpolation, but in this paper we have not done so.  As
a result, the reconstructed sources have a fragmented appearance on
small scales, as we will see later in
Figs.~\ref{fig:match:J0029}--\ref{fig:match:J1627}.

While the fitting of synthetic images is essentially the same as in conventional lensing modelling
\citep[our implementation is the same as in][]{Denzel20, Denzel20b},
plausible-matching requires a further issue to be solved, namely the alignment
and orientation of the lens system.  The nuisance parameters $\xi = (p_{\rm
rel}, \phi_{\rm rel})$, where $p_{\rm rel}$ is the position, $\phi_{\rm rel}$ is
the orientation of the mass map relative to the observation, needs to be
marginalised out.  The marginalisation is done using short Markov-Chain
Monte-Carlo (MCMC) simulations.  The result is an ensemble of plausible-matches,
reminiscent of model ensembles in free-form lens modelling
\citep{PixeLens,GLASS} but having a different meaning, because they arise from
galaxy-formation simulations.

The minimum of $\chi^2$ in \eqref{match:synthetics_chisqr} need not
correspond to a unique $\bm\alpha$.  In other words, very different
galaxies can in principle produce identical synthetic images.  This is
the well-known problem of lensing degeneracies \citep[for a review,
  see][]{Wagner18}.  The plausible matching strategy automatically
marginalises over simulated galaxies that are degenerate in the
observables, so lensing degeneracies as such are not an obstacle to
the method.  If, however, the differences between galaxy-formation
scenarios happen to be aligned along lensing degeneracies, lensing
observables would be ineffective as discriminators between
galaxy-formation models.  Such a thing seems unlikely, but we cannot
rule it out at present.

In total, 11634 MCMC simulations had to be executed until the solutions for all
simulated galaxies and lens systems converged.  This was relatively easily
achieved within about 4--8 hours per lens through some optimisations and
some compromises.  The inclusion of a PSF increases the
non-sparseness of the synthetic-image mapping considerably, makes the generation
of synthetics quite computationally intensive, and slows down the MCMC
simulations by an average factor of $\sim50$.  Fortunately, initial tests showed
that the omission of the PSF in \eqref{match:mapping} for this step caused
acceptable differences.  Since both the projected surface-density maps and
cutouts from the observations have been centred well beforehand, $p_{\rm rel}$
never deviated from the centre by more than $0.05\arcsec$, which lead us to
discard that parameter in the final stage.  The convergence to optimal alignment
rotation angles on the other hand was more relevant, especially for galaxies
with high ellipticity, whereas for round galaxies the rotation angles were
arbitrary and usually settled around $0\degree$.

The subsequently described lens-matching method has been implemented in the
public software \Code{gleam}\footnote{\faGithub\,
\url{https://github.com/phdenzel/gleam}} (Gravitational Lens Extended Analysis
Module) by PD.  It is written in Python and thus comes with all of its
flexibility and a large scientific library support.  Computationally demanding
tasks such as the calculation of potential gradients are alternatively also
implemented in a mixture of C and Cython \citep{cython}.  Similar to the lens
modelling tool \Code{GLASS}\footnote{\faGithub\,
\url{https://github.com/jpcoles/glass}} by \cite{GLASS}, the module encompasses
more general features, some of which are still in development, but the
lens-matching technique lies at its core.  In particular, the synthetic imager
described in \subsecref{match:synthetics} is implemented in the sub-module
\Code{gleam.reconsrc}.

The entire lens-matching method was intentionally kept relatively
simple and lightweight in order to keep it scalable for a much bigger
lens sample using larger catalogues and to minimize the input required
from the outside.  \figref{match:schematics} shows a schematic graph
which summarizes each key step of the lens-matching method.  The
analysis presented here aimed for a proof-of-concept only.  With a
working basis, further refinements and improvements can easily be
explored in isolation and afterwards properly implemented.  In
\secref{match:conclusion}, we give some suggestions of what aspects
could be improved first.

\begin{figure}%
    \centering%
    \includegraphics[width=0.475\textwidth]{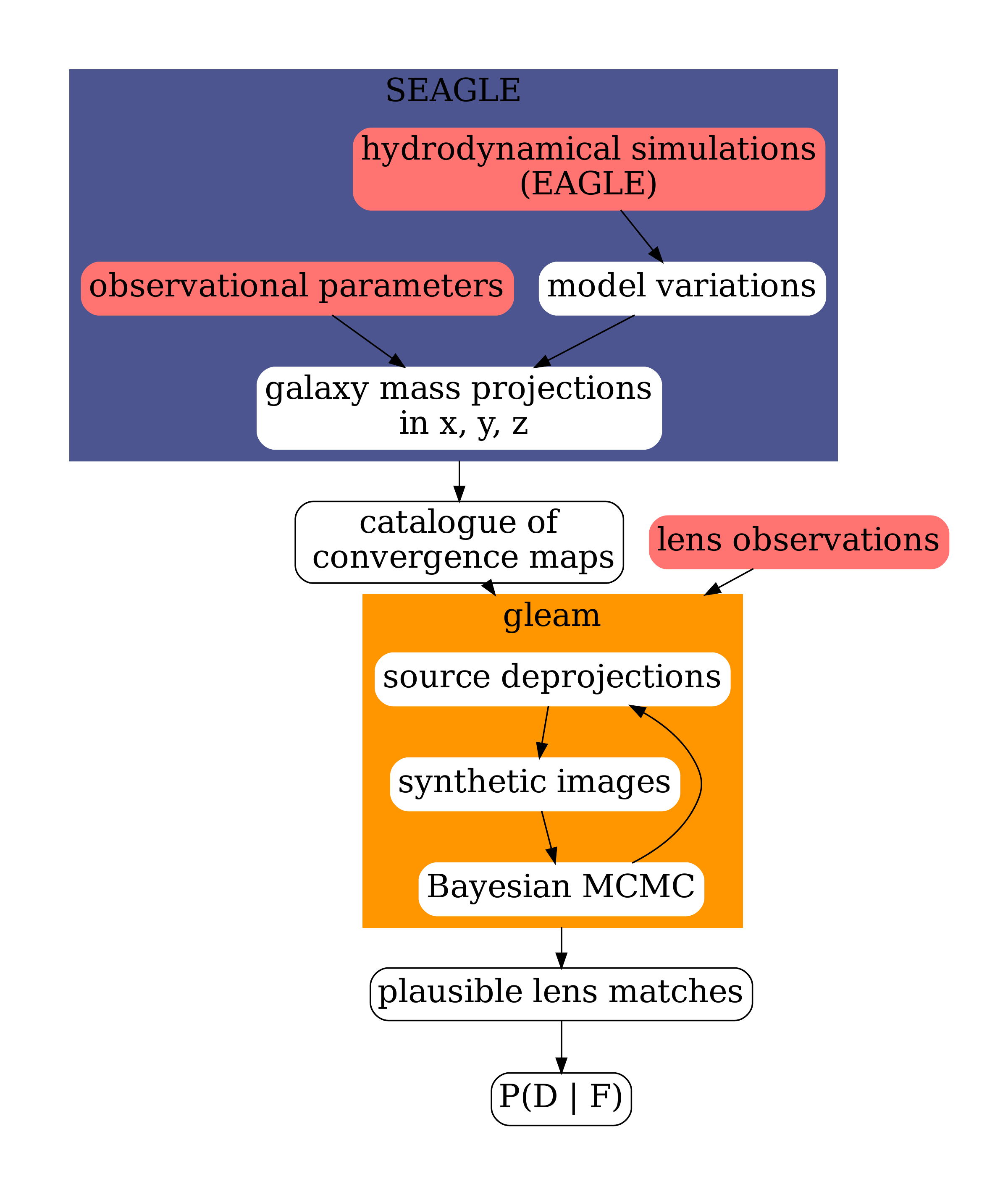}%
    \caption{A schematic top-bottom graph summarizing the specific
      steps of the lens-matching method in this work.  The pipeline
      was intentionally kept quite modular; in particular, the
      components in the blue and yellow coloured subgraphs are
      specific choices of this work, and could in principle be
      replaced by equivalent operators.  Red fields represent
      independent data inputs from observations or simulations.
    }\figlbl{match:schematics}%
\end{figure}%

\section{SEAGLE}\seclbl{match:seagle}

\cite{SEAGLE-I} introduced the \Code{SEAGLE} pipeline to systematically study
galaxy formation via simulated strong lenses from the EAGLE simulations
\citep{EAGLE, Crain15, McAlpine16}. \Code{SEAGLE} used the \Code{GLAMER}
ray-tracing package \citep[Gravitational Lensing with Adaptive Mesh Refinement;
][]{Metcalf14, Petkova14} to create realistic lensed images and calculate all
other lensing quantities used in their analysis.  \Code{SEAGLE} aims to
investigate and possibly disentangle galaxy formation and evolution mechanisms
by creating, modelling, and analysing simulated strong lens-galaxies to compare
them with observations.

EAGLE is a suite of state-of-the-art hydrodynamical simulations that explored
several feedback scenarios and model variations giving us a set of galaxy
evolution scenarios to assess their impact on the present-day universe.
\cite{Crain15} divided the simulations into two categories. The first comprises
four simulations {\sl calibrated} to yield the $z=0.1$ galaxy stellar mass
function (GSMF) and central black hole (BH) masses as a function of galaxy
stellar mass.  The second category comprises simulations that each vary a single
sub-grid physics parameter with respect to the Reference model but without
considering whether they match the GSMF (i.e.\ they are not calibrated).

In \citet{SEAGLE-II}, using \Code{SEAGLE}, the authors quantified that if the
simulated lensed images are modelled similar to the observations, then the
median total mass density slope of galaxies from an inefficient AGN feedback
model (AGNdT8: Reference variation) and a constant feedback model (FBconst:
Calibrated simulation) that becomes inefficient at denser environment gives
slopes $t$=2.01 and  $t$=2.07, respectively, in good agreement with the
observations of SLACS, SL2S (Strong Lensing Legacy Survey), and BELLS (Baryon
Oscillation Spectroscopic Survey (BOSS) Emission-Line Lens Survey). Galaxies in
the EAGLE Reference model (benchmark model), however, tend to have a steeper
median total mass density slope ($t$=2.24) than observed lens galaxies (i.e.\
$t$ =2.08 for SLACS, $t$=2.11 for BELLS and $t$=2.18 for SL2S).

The nomenclature of the \Code{SEAGLE}-projected mass distributions in the
catalogue depends on their halo, subhalo, and projection axis.  A number
following 'H' refers to the halo number, 'S' gives the subhalo, and letters
'A/B/G' refers to the projection the galaxy has undergone in Cartesian
coordinates i.e. $\alpha$, $\beta$ and $\gamma$ respectively.  The feedback
model denominations are prepended in this nomenclature.

For our analysis, we choose these two galaxy evolution scenarios (AGNdT8 and
FBconst) as they are most realistic to the strong lensing observations. We
briefly discuss the key features of these feedback models below.

In the calibrated simulations, the models differ in terms of their adopted
efficiency of feedback associated with star formation, and how this efficiency
depends upon the local environment. The general consensus shows that the
properties of simulated galaxies are most sensitive to the efficiency of
baryonic feedback \citep[see e.g.,][]{Schaye10, Vogelsberger13}. Below a certain
resolution limit, the physical processes cannot be simulated via the dynamics of
the particles. So they are incorporated via analytic prescriptions in all
hydro-dynamic simulations including EAGLE. In EAGLE model variations, the
efficiency of the stellar feedback and the BH accretion were calibrated to
broadly match the observed local ($z \approx$ 0) GSMF.  Also, several studies
established that AGN feedback is a necessary ingredient for regulating the
growth of massive galaxies \citep[e.g.][]{Crain09, Schaye10, Haas13a}.

Below we briefly describe the EAGLE galaxy formation models which were used in
this work.

\subsection{A Constant Feedback (FBconst)}\subseclbl{match:FBconst}

The simplest feedback model used in EAGLE is FBconst. In this calibrated model,
independently from the local conditions, a fixed amount of energy per unit
stellar mass is injected into the ISM. This fixed energy corresponds to the
total energy discharged by type-II SNe ($f_{\rm th}=1$). While the stellar
feedback in this model was not calibrated, the model does reproduce the
observables used for the calibration.  \citet{Crain15} found that the thermal
stellar feedback prescription employed in EAGLE becomes inefficient at high gas
densities due to resolution effects (\citealt{Vecchia12}). Thus in this model,
there is a lack of compensation for more energy at higher gas density. Thus the
stellar feedback will be less effective in high-mass galaxies (where the gas
tends to have higher densities) (\citealt{Crain15}).

\citet{EAGLE} demonstrated that it is possible to calibrate the Reference model
to reproduce the GSMF and the observed sizes (in different bands) of galaxies at
$z$ = 0.1. \citet{Crain15} conducted a series of simulations (listed in the
lower section of Table 1 therein) for which the value of a single parameter was
varied from that adopted in the Reference model. One of the parameters varied
was AGN temperature. 

\subsection{Temperature variation in AGN heating (AGNdT8)}\subseclbl{match:AGNdT8}

\cite{EAGLE} have examined the role of the AGN heating temperature in EAGLE by
adopting $\Delta T_{\rm AGN}=10^{8.5}{\rm K\; and\;}10^9 {\rm K}$. They
demonstrated that a higher heating temperature produces less frequent but more
energetic AGN feedback episodes. They concluded it is necessary to reproduce the
gas fractions and X-ray luminosities of galaxy groups. \citet{LeBrun14} also
concluded that higher heating temperature yields more efficient AGN feedback.
There are two Reference-model variation simulations with $\Delta T_{\rm
AGN}=10^{8} \rm K$  (AGNdT8) and $\Delta T_{\rm AGN}=10^{9} \rm K$  (AGNdT9),
besides the Reference model itself which adopted $\Delta T_{\rm AGN}=10^{8.5}
\rm K$. In massive galaxies, the heating events (less frequent but more
energetic) are more effective at regulating star formation due to a higher
heating temperature. AGNdT8 (AGNdT9) model has a higher (lower) peak star
fraction compared to the Reference model. The reduced efficiency of AGN
feedback, when a lower heating temperature is adopted, leads to the formation of
more compact galaxies because gas can more easily accrete onto the centers of
galaxies and form stars. \citet{SEAGLE-I} showed that for galaxy-galaxy strong
lenses, AGNdT8 produces closest analogs for SLACS.  Thus, for this work, we use
galaxies from the AGNdT8 simulation, in addition to the galaxies from the
simpler FBconst model.

\section{The observed systems}\seclbl{match:testcases}

In order to test whether searching for plausible matches from EAGLE simulations
is at all feasible, we selected a small sample of seven lens systems that have
already been studied by other methods.  The selection was based on three
criteria.  First, the system had to be clearly strongly lensed, with relatively
easily identifiable images showing very clear evidence of multiple imaging.
Second, the observations needed to have extended images and arcs with some
imperfections (rather than point-like lensed quasars) so as to challenge the
matching technique.  Third, the sample had to be representative of a larger
sample of lenses.  The third criterion made it natural to choose from SLACS, and
from the SLACS lenses of quality category “A” we selected seven, having a wide
range of mean image radii, which is a rough proxy for Einstein radii.
Figs.~\ref{fig:match:J0029}--\ref{fig:match:J1627} in their top left panels show
the lensed images in HST-image F814W bands.  The most relevant information about
the systems is listed in \tabref{match:systems}, including references to the
discovery papers.

\begin{table*}
    \centering
    \newcommand{\tref}[1]{\hyperlink{tr#1}{(#1)}}
\newcommand{\tlink}[1]{\hypertarget{tr#1}{(#1)}}

\begin{tabular}{lrlccrcl}
        \hline\hline
        Name & R.A. [hms] & Decl. [dms] & z$_{\mathsf{l}}$ & z$_{\mathsf{s}}$ & $\sigma_{\mathsf{SDSS}}$ [km/s] & R$_{\mathsf{eff}}$ [arcsec] & Reference \\
        \hline
        SDSSJ0029$-$0055 & 00:29:07.8 & $-$00:55:50 & 0.23 & 0.93 & 229 $\pm$ 18 & 2.16 & \tref{1} \\
        SDSSJ0737$+$3216 & 07:37:28.5 & $+$32:16:18 & 0.32 & 0.58 & 310 $\pm$ 15 & 2.16 & \tref{2} \\
        SDSSJ0753$+$3416 & 07:53:46.2 & $+$34:16:33 & 0.14 & 0.96 & 208 $\pm$ 12 & 1.89 & \tref{3} \\
        SDSSJ0956$+$5100 & 09:56:29.8 & $+$51:00:06 & 0.24 & 0.47 & 299 $\pm$ 16 & 2.33 & \tref{2} \\
        SDSSJ1051$+$4439 & 10:51:09.4 & $+$44:39:08 & 0.16 & 0.54 & 216 $\pm$ 16 & 1.66 & \tref{3} \\
        SDSSJ1430$+$6104 & 14:30:34.8 & $+$61:04:04 & 0.17 & 0.65 & 180 $\pm$ 15 & 2.24 & \tref{3} \\
        SDSSJ1627$-$0053 & 16:27:46.5 & $-$00:53:57 & 0.21 & 0.52 & 275 $\pm$ 12 & 2.08 & \tref{2} \\
        \hline
\end{tabular}
    \caption{\tablbl{match:systems}%
    Summary of the observed lensing systems.
    \protect\tlink{1} \protect\cite{SLACS5};
    \protect\tlink{2} \protect\cite{SLACS1};
    \protect\tlink{3} \protect\cite{SLACS13}}%
\end{table*}%

\subsection{SDSSJ0029\textminus0055}\subseclbl{match:SDSSJ0029}

SDSS\SDSSJ{0029}{} appears to be a relatively small, doubly lensing system
observed on 12 September 2006.  Initial reports by \cite{SLACS5} classify it as
a single, early-type galaxy.  The redshift of the foreground galaxy was
spectroscopically measured to $z_L = 0.2270$ and for the background source to
$z_S = 0.9313$.  It exhibits an almost fully closed ring which is relatively
difficult to recognize due to the light pollution from the foreground galaxy.  A
de~Vaucouleurs fit to the lensing galaxy gives a relatively high effective
radius of 2.16\arcsec.  It also has a well measured stellar velocity dispersion
of $\sigma_{\text{SDSS}} = 229 \pm 18\;\kms$.

The initial report presented a singular isothermal ellipsoid and
light-traces-mass gravitational lens model which provided best fits using two
source-plane components.  However, the present work indicates that a
single-component source (see second image in the left column in
\figref{match:J0029}) is also possible.  The top image in \figref{match:J0029}
in the left column shows the system from the HST/ACS-WFC1 observation (Advanced
Camera System Wide Field Channel 1) using the F814W filter.

\subsection{SDSSJ0737+3216}\subseclbl{match:SDSSJ0737}

SDSS\SDSSJ{0737}{} appeared in the first SLACS report by \cite{SLACS1}.  A
successive report \citep{SLACS5} grades the quality of the single-multiplicity,
early-type galaxy to be of type ``A''.  Its foreground and background redshifts
were measured to $z_L = 0.3223$ and $z_S = 0.5812$ respectively.  From SDSS
spectroscopic data, a good estimate for the velocity dispersion of
$\sigma_{\text{SDSS}} = 338 \pm 17\;\kms$ was provided.

Parametric lens models from \cite{SLACS5} used two source-plane components to
fit the astrometric data.  The HST/ACS-WFC1 observation (on 21 September 2004)
using the F814W filter is displayed in the top image in the left column of
\figref{match:J0737}.  It shows two extended images of which the brighter image
is most likely the product of two merged ones, and a point-like image connected
via a dim arc, which would most likely classify it as a short-axis quad.
Besides the initial modelling, \cite{Ferreras07} provided a free-form lens model
for this lens system, along with a spatially resolved comparison to the
stellar-mass surface distribution derived from population-synthesis models.

\subsection{SDSSJ0753+3416}\subseclbl{match:SDSSJ0753}

The top image of the left column in \figref{match:J0753} shows
SDSS\SDSSJ{0753}{} (HST/ACS WFC1 F814W on 8 September 2010) as a clearly lensing
system. It is a very interesting system with minimum eight (possibly even 12)
lensed images of at least two sources.  \cite{SLACS13} reports ellipsoid lens
models using even four source-plane components.  Either way, this lens promises
a much lower degree of degeneracy due to the high number of lensed images and
sources.

From the SDSS observations the lensing galaxy was classified as an early-type,
single-multiplicity foreground galaxy with a well measured velocity dispersion
$\sigma_{\text{SDSS}} = 206 \pm 11\;\kms$.  The redshift estimates for the lens
and source are $z_L = 0.1371$ and $z_S = 0.9628$, respectively.

\subsection{SDSSJ0956+5100}\subseclbl{match:SDSSJ0956}

In the left column of \figref{match:J0956}, the top image shows
SDSS\SDSSJ{0956}{} from the HST/ACS-WFC1 F814W observation from 1 November 2006.
\cite{SLACS1} designates its lens an early-type, single foreground galaxy.  The
spectroscopic survey yielded a velocity dispersion of $\sigma_{\text{SDSS}} =
299 \pm 16\;\kms$ and redshifts of $z_L = 0.2405$ and $z_S = 0.4700$ for the
foreground and background source respectively.

The lens shows four lensed source images at various angular separations from the
lensing galaxy in a short-axis quad configuration.  Two of the images appear
rather point-like and bright whereas the others are extended and fainter.

It is another lens which was free-from modelled by \cite{Ferreras07}.  The
models predict it to be a rather high-mass galaxy with a total and stellar mass
within the aperture image radius $R_M$ of 
\begin{equation*}
    \begin{aligned}
    &M_{\text{tot}}(<R_M) = 66.4^{+25.7}_{-16.7} \cdot 10^{10}\,M_{\odot}\ \text{and}\\
    &M_{\text{star}}(<R_M) = 41.8^{+4.9}_{-4.0} \cdot 10^{10}\,M_{\odot},
    \end{aligned}
\end{equation*}
where the aperture radius is $2R_{\rm max} - R_{\rm min}$, the
difference of projected radii of twice the outermost and innermost
lens images.

\subsection{SDSSJ1051+4439}\subseclbl{match:SDSSJ1051}

\cite{SLACS13} reports SDSS\SDSSJ{1051}{} as another early-type, single
foreground-galaxy lens system.  The SDSS data yields a velocity dispersion value
of $\sigma_{\text{SDSS}} = 216 \pm 16\;\kms$;  the redshifts of the lens and
background source are reported with $z_L = 0.1634$ and $z_S = 0.5380$
respectively.  While the image shown in \figref{match:J1051} (left column, top
panel) does not clearly indicate that the supposedly doubly lensing system is
indeed lensing, its type is still classified as ``A''.

\subsection{SDSSJ1430+6104}\subseclbl{match:SDSSJ1430}

\figref{match:J1430} (first image in the left column) depicts SDSS\SDSSJ{1430}{}
(HST/ACS-WFC1 F813W) as a very noisy lens system with faint lensed images, with
considerable pollution by the host galaxy.  \cite{SLACS13} reports the
early-type galaxy with a velocity dispersion value of $\sigma_{\text{SDSS}} =
180 \pm 15\;\kms$.  The SDSS redshifts for the foreground and background objects
are $z_L = 0.1688$ and $z_S = 0.6537$.

The parametric, ellipsoid lens models use two source-plane components to fit the
photometric data, with a total mass within the Einstein radius of $1.02 \cdot
10^{11}\,\mathrm{M}_{\odot}$.

\subsection{SDSSJ1627\textminus0053}\subseclbl{match:SDSSJ1627}

On 12 March 2006, the HST/ACS-WFC1 observed SDSS\SDSSJ{1627}{} as a double with
an almost completely closed ring.  \cite{SLACS1} reported it as an early-type
foreground galaxy in a single-multiplicity system with redshifts $z_L = 0.2076$
and $z_S = 0.5241$.  The spectroscopically determined velocity dispersion is
well measured with a value of $\sigma_{\text{SDSS}} = 290 \pm 15\;\kms$.  A
picture of the lens system can be found on the top panel of the left column in
\figref{match:J1627}.

\section{Results}\seclbl{match:results}

The results on plausible matches for the seven lens systems considered are
displayed in Figs.~\ref{fig:match:J0029}--\ref{fig:match:J1627} and summarised
in \tabref{match:models}. Figs.~\ref{fig:match:J0029}--\ref{fig:match:J1627} are
devoted to one lens each, in the same order as in \tabref{match:systems}.

Each of these figures has eight panels, arranged as
follows.
\begin{center}
\begin{tabular}{cc}
Observed image & Synthetic image \\
Source brightness & Residual Image \\
Lensing mass & Pixelized lensing mass \\
Lensing Roche potential & Mean enclosed convergence \\
\end{tabular}
\end{center}

We now discuss properties of the most-plausible matches as shown in
Figs.~\ref{fig:match:J0029}--\ref{fig:match:J1627}.

\subsection{Images}\subseclbl{match:synths}

The top row of Figs.~\ref{fig:match:J0029}--\ref{fig:match:J1627} shows the
observed lensed images and the synthetic image from the most-plausible match.
The lensing galaxy is masked out.  The difference between these, scaled by the
noise --- in other words, the pixelwise $\chi^2$ from
\eqref{match:synthetics_chisqr} --- is shown in the right panel of the second
row.

For the MCMC over the orientation $\phi_{\rm rel}$ it is computationally simpler
to rotate the image rather than the lens.  As a result, there are some
rotated-corner artifacts, less noticeable in the synthetic images, but at the
edges of the source plane, especially for example in \figref{match:J1051}.
These are, however, harmless $\chi^2$ computation, for which only a circular
region was considered.

In all of the second-row right panels, it is evident that the contributions to
$\chi^2$ come mainly from an annular region where the multiply-imaged features
are.  The black inner disc is of course just the masked-out lensing galaxy.  The
outer part in the $\chi$ maps is dark (or at a lower level) because without
multiple images the source brightness has the trivial solution
\begin{equation}\eqlbl{match:trivial}
s(\bm\theta-\bm\alpha(\bm\theta))=I^{\rm obs}(\bm\theta)
\end{equation}
and then any contribution to $\chi^2$ comes only because there are fewer pixels
in the source plane than in the image plane.  It would be better to consider
only the multiply-imaged region when computing $\chi^2$, but it is not clear how
to do so efficiently.

\subsection{Source reconstructions}\subseclbl{match:srcs}

The left panel in the second row in each of
Figs.~\ref{fig:match:J0029}--\ref{fig:match:J1627} shows the
reconstructed $s(\bm\beta)$ for the most-plausible match.
The sources appear fragmented on small scales because of a discretisation
artefact explained in \subsecref{match:synthetics} which we have
not interpolated out.

The source-fitting as implemented here does not guarantee that the reconstructed
source will be blob-like and not a random scatter plot.  However, plausible
matching lenses are generally associated with plausible looking source maps (disregarding the small-scale fragmentation).  In
cases where the data are more noisy the source plane also tends to be noisy;
this is especially noticeable in \SDSSJ{0737}, \SDSSJ{1051}, and \SDSSJ{1430}\
(Figures \figref*{match:J0737}, \figref*{match:J1051}, and
\figref*{match:J1430}), where the bright specks from the data images (probably
artifacts left by cosmic particles) also appear in the source plane.  While most
sources seem to be rather symmetric, \SDSSJ{0753}\ (\figref{match:J0753}) and
\SDSSJ{0956}\ (\figref{match:J0956}) appear to be amorphous with multiple cores.
This could be an indication of a merging system or multi-component source,
however further investigation is needed to confirm this.

A curious artifact appears in the cases of \SDSSJ{0956}\ (\figref{match:J0956})
and \SDSSJ{1051}\ (\figref{match:J1051}).  There the source appears to have
bright edges in a curved diamond shape.  The curved edges evidently correspond
to the diamond caustic for four-image lenses, which correspond extreme
magnification, and single pixels along these edges can map to large areas on the
image plane.  We conjecture that the source-fitting procedure is using this
property of caustics to fit noise in the images.

In comparison with source reconstructions from previous works, some differences
are noticeable. In most cases, general shapes of the sources agree with previous
works, when noise is ignored, especially for \SDSSJ{0029}. For \SDSSJ{0753}\ and
\SDSSJ{1430}\ the main cores exhibit similar shapes, but previous works include
more secondary sources compared to most source reconstructions here. Contrarily,
\SDSSJ{0956}, although being very noisy, seems to exhibit more components than
reconstructions from previous works.

\subsection{Mass maps}\subseclbl{match:kappa}

The third row in each of Figs.~\ref{fig:match:J0029}--\ref{fig:match:J1627}
shows the $\kappa$ maps from \Code{SEAGLE} and the reduced-resolution $\kappa_n$
maps that we actually used, for the most-plausible match. The dark contours
indicate $\kappa=1$.

Interestingly, while the catalogue did include many projected surface-densities
with high ellipticity, the lens-matching approach seems to preferentially select
rather round models.  However, this of course depends on the selection of the
lens system and considering to the light profiles of the lenses in the data,
mass distributions with low ellipticity were to be expected.  The mass models
do, however, exhibit a moderate amount of substructure.

The bottom-right panel in each of
Figs.~\ref{fig:match:J0029}--\ref{fig:match:J1627} shows the mean enclosed
density $\langle\kappa\rangle_\theta$ within a given angular radius for the 10
most plausible matches in the sense of $\chi^2$.  As in \figref{match:catalogue}
$\langle\kappa\rangle_\theta=1$ is understood as the Einstein radius.  The value
is well-constrained, even if we consider the 50 most-plausible matches as
illustrated, or in a subset of best-matching models with $\chi^{2}_{\nu} < 5$
as in \tabref{match:models}.

\subsection{Lensing Roche potentials}\subseclbl{match:arriv}

The bottom-left panels in Figs.~\ref{fig:match:J0029}--\ref{fig:match:J1627}
show another interesting quantity, a contour map of the \textit{lensing Roche
potential}
\begin{equation}
\roche(\bm\theta) = {\textstyle\frac12}\theta^{2} - \psi(\bm\theta)
\end{equation}
which we introduced in \cite{Denzel20}.  The lens
equation~\eqref*{match:lens_equation} is equivalent to
\begin{equation}
\bm\beta = \nabla\roche(\bm\theta)
\end{equation}
and consequently the points where $\nabla\mathcal{P}=0$ are image locations from
a source at $\bm\beta=0$.  These points are extrema (minima, maxima, and
saddle-points) of $\nabla\mathcal{P}$ and easy to discern on a contour map.  The
actual image positions will be somewhat different, depending on the details of
$s(\bm\beta)$, but nevertheless, the contours of the lensing Roche potential
offer a simple confirmation that a plausible match is indeed a strongly lensing
system, and that we have not simply stumbled upon the trivial solution
\eqref*{match:trivial}.

\subsection{Relative posteriors}\subseclbl{match:posteriors}

Every pre-selected model was match-tested against the
observational data of each lens according to \eqref{match:synthetics_chisqr},
which yielded distributions of reduced least squares $\chi^{2}_{\nu}$.
\figref{match:chi2_mosaic} shows these distributions as cumulative histograms,
including the fractions of models from the two galaxy-formation scenarios,
FBconst and AGNdT8. 

Subsets of most-plausible matches, that is, matches with minimal $\chi^{2}_{\nu}$,
are likely to contain models from both galaxy-formation scenarios, evident in
\tabref{match:models} and \figref{match:chi2_mosaic}.

\begin{table*}
    \centering \begin{tabular}{lllllrrr}
\hline\hline
Lens & $N_{\chi^{2}_{\nu} < 5}$ & $N_{\chi^{2}_{\nu} < 5}$ & $\chi^{2}_{\nu}$ & $\chi^{2}_{\nu}$ & $\delta\phi_{\mathrm{rel}}$ & $\theta_{\mathrm{E}}$ & $M_{\mathrm{stel}}$\\
 & AGNdT8 & FBconst & AGNdT8 & FBconst & [\degree] & [\arcsec] & [$10^{11}\,M_{\odot}$] \\ \hline
 SDSSJ0029$-$0055 &  52 &  86 & 2.71 & 2.68 &  11.1$^{+3.3}_{-3.3}$ &  1.04$^{+0.08}_{-0.08}$ &  1.30$^{+0.20}_{-0.22}$ \\
 SDSSJ0737$+$3216 &   6 &   2 & 3.74 & 3.47 &   4.8$^{+0.7}_{-0.7}$ &  1.09$^{+0.01}_{-0.03}$ &  3.84$^{+1.56}_{-0.88}$ \\
 SDSSJ0753$+$3416 &  61 &  48 & 2.78 & 2.84 &  15.8$^{+3.1}_{-3.1}$ &  1.39$^{+0.11}_{-0.09}$ &  1.67$^{+0.21}_{-0.53}$ \\
 SDSSJ0956$+$5100 &   4 &   6 & 3.50 & 3.68 &   3.7$^{+1.0}_{-1.0}$ &  1.48$^{+0.16}_{-0.03}$ &  5.85$^{+1.17}_{-0.19}$ \\
 SDSSJ1051$+$4439 &  17 &  24 & 2.90 & 2.69 &   9.4$^{+2.1}_{-2.1}$ &  1.56$^{+0.16}_{-0.09}$ &  3.53$^{+0.66}_{-0.40}$ \\
 SDSSJ1430$+$6104 &  41 &  58 & 2.49 & 2.65 &   5.1$^{+1.2}_{-1.2}$ &  1.22$^{+0.07}_{-0.11}$ &  1.84$^{+0.20}_{-0.35}$ \\
 SDSSJ1627$-$0053 &  30 &  33 & 2.37 & 2.48 &  17.5$^{+5.5}_{-5.5}$ &  1.40$^{+0.12}_{-0.08}$ &  3.52$^{+0.33}_{-0.50}$ \\
\hline
\end{tabular}

    \caption{\tablbl{match:models} Results for a subset of the most
      plausible matches for each lens system with $\chi^{2}_{\nu} < 5$.
      The reduced $\chi^{2}_{\nu}$ apply to the best synthetic images of the
      matching tests, $\delta\phi_{\mathrm{rel}}$ are the average deviations in
      orientations about the line of sight of the 68\% interval from the MCMC of
      all models in the subset. The $\theta_{\mathrm{E}}$ column contains
      medians of Einstein radii and the $M_{\mathrm{stel}}$ column
      medians of the total mass in stars of the simulated galaxy models, with
      uncertainties covering the 68\% interval of the model subset.}
\end{table*}

\begin{figure}%
    \centering%
    \includegraphics[width=0.475\textwidth]{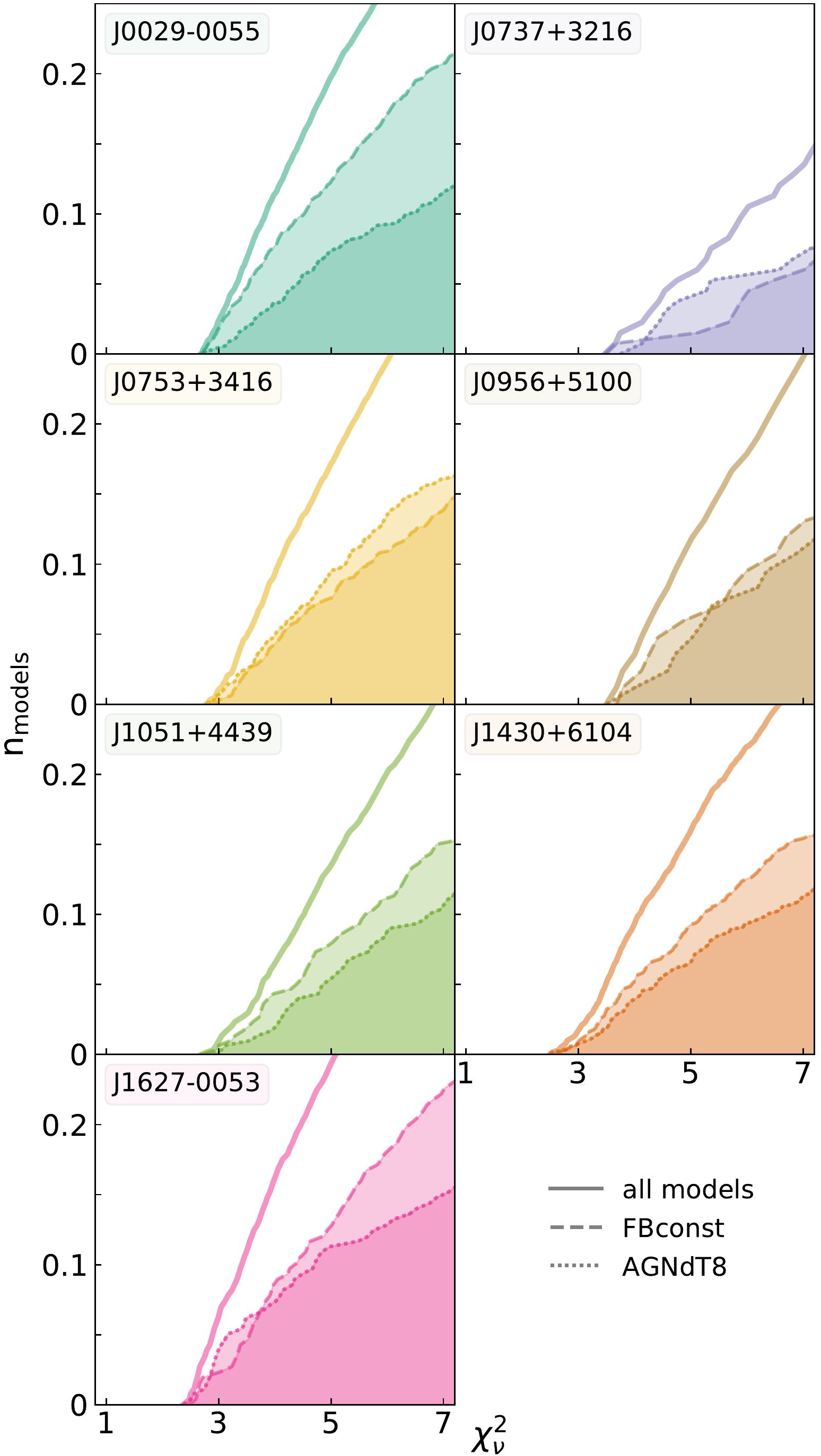}%
    \caption{ Cumulative $\chi^{2}_{\nu}$ histograms for all match-tested
      models.  The dotted lines show the fractions of models with
      feedback scheme AGNdT8, whereas the dashed lines are for FBconst
      models. The numbers of models $n_{\mathrm{models}}$ were
      normalized by the total numbers of models, preselected
      for each lens individually. Models with high $\chi^{2}_{\nu}$ are
      considered bad matches to the observations. For instance, for
      the lens systems \protect\SDSSJ{0737}\ and \protect\SDSSJ{0956}\
      only a few plausible matches have been found within the model
      catalogue.  }\figlbl{match:chi2_mosaic}%
\end{figure}%

Considering the most-plausible matches with e.g. $\chi^{2}_{\nu} < 5$,
we can evaluate a Bayesian evidence in the form
$$ \frac{P(\hbox{AGNdT8}\,|\,D)}{P(\hbox{FBconst}\,|\,D)} $$ for each lens.
Using the values from \tabref{match:models}, this is between 0.6 and 0.7 for
\SDSSJ{0029}, \SDSSJ{0956}, \SDSSJ{1051}, and \SDSSJ{1430}, meaning these
systems show a slight tendency towards FBconst.  For \SDSSJ{0737}\ and
\SDSSJ{0753}\ the expression above evaluates to well above 1.0, indicating a
tendency towards AGNdT8, whereas for \SDSSJ{1627}\ it is very close to 1.0.
Although, it should be noted that these values are not significant yet
(especially for \SDSSJ{0737}\ and \SDSSJ{0956}) and for better statistics more
matching tests should be performed.  With more match-tests and better
statistics, the criterion $\chi^{2}_{\nu} < 5$ can be lowered to ensure that
only the best-matching models are used to evaluate the relative posterior
probability distribution. This threshold depends on the individual systems and
galaxy models used in the matching method. In this case, models with
$\chi^{2}_{\nu}$ above 5 start to display various noticeable deficits in the
source reconstructions and synthetic images and are therefore not suitable to
estimate the relative posterior.

\tabref{match:models} also lists the subsets' median values of the Einstein
radii which is a measurement of the total mass the lens. These values are
consistent with previous studies \citep{SLACS5, SLACS1, SLACS13, Ferreras07}.
The comparison of the median stellar masses of the model galaxies with previous
estimates also seem to agree well, if it is considered that previous estimates
are within an Einstein radius or half-light radius of the lensing galaxies,
whereas for our models it is possible to estimate the entire mass in stars.

\section{Discussion}\seclbl{match:conclusion}

Mass reconstructions in gravitational lensing are in general non-unique.  Even
for strong-lensing clusters with tens of multiply-imaged systems over a range of
redshifts, there is significant scatter among mass models even if they fit the
data equally well \citep[see e.g.,][]{FrontierFieldsTest}.  For galaxy lenses
the non-uniqueness of models is much more evident, and indeed has been known
since the earliest days of lens modelling \citep{Young1981b}.  This facts
suggests that the large catalogues of simulated galaxies in recent
galaxy-formation simulations may contain plausible matches to individual
observed lensing galaxies.  In this work we search for and find plausible
matches among EAGLE simulated galaxies to seven observed lensing galaxies from
SLACS.  The main computational part is to fit for (a)~an orientation of a given
simulated galaxy and (b)~a source light distribution, such that the observed
light distribution is reproduced.  This is implemented in the new \Code{gleam}
code, but automated lens-modelling tools such as \Code{AutoLens}
\citep{AutoLens} and \Code{Ensai} \citep{Hezaveh17} could probably also be
adapted for the purpose, if required.

The main conclusion of this work is that EAGLE --- and presumably
other comparable galaxy-formation simulations --- contain plausible
matches for observed lensing galaxies.  Hence it appears feasible to
use observed lensing galaxies as constraints on galaxy-formation
scenarios, without conventional lens models.  Obtaining statistically
significant results, however, will need several issues to be addressed
first, which we discuss briefly below.

\begin{enumerate}
\item In this work we have used single simulated galaxies,
  disregarding the environment and line-of-sight structures, and also
  approximated the projected mass as consisting of $23\times23$ mass
  tiles.  Furthermore, we have considered rotations only about three
  orthogonal candidate lines of sight, rather than arbitrary
  orientations in 3D.  All these aspects of the implementation need to
  be improved, while keeping it efficient for the purpose of scaling
  up to larger lens samples.

\item The source reconstruction is another area that can be improved.  The
  advantage of the procedure used in this work is that supervision at the level
  of individual lenses is not required, though this will not be true if the
  observation data is dominated by noise and extraneous light, because
  additional masks would be needed.  The disadvantage of the current procedure
  is that the fitted source is just an arbitrary brightness map, and the
  principle of plausible matches is not being applied.

\item Since plausible-matching galaxies for any given lens always have very
  similar Einstein radii, even though they may differ in other ways, it is
  advantageous to pre-select the simulated galaxies to be within a suitable
  range of Einstein radii.  In this work, we produced a conventional lens model
  first, but a more efficient method is desirable.

\item Provided the lensing galaxy is clearly visible in the data, stellar mass
  estimated from multi-band images using population synthesis
  \citep[cf.][]{Leier16} could be incorporated into the likelihood $P(D\,|\,g)$.
  Stellar-maps from the simulations are, of course, known a priori.  Ideally,
  the stellar light distribution would be subtracted from the entire
  observational data using models of the galaxy light.

\item Stellar kinematics would be an important ingredient in $P(D\,|\,g)$.
  Current simulations soften the gravitational dynamics on scales of order a kpc
  \citep[see e.g., Table~2 in][]{EAGLE}, and it would be interesting to see if
  this strongly affects $P(D\,|\,g)$.  It would also be interesting to see if an
  equilibrium galaxy-modelling framework like \Code{AGAMA}, which resolves much
  smaller scales, yields higher $P(D\,|\,g)$ than cosmological simulation.

\item Finally, although available for only a small fraction of lenses, time
  time-delays \citep[for recent observations see][]{Millon20} would be
  interesting to incorporate in the plausible-match scheme.  Lensing time delays
  are usually thought of as a way of measuring cosmological parameters,
  especially $H_0$.  But the accuracy of the $H_0$ inferences from lensing
  depends on how well $P(g\,|\,F)$ of the universe is constrained.  Hence time
  delays could be useful (if they turn out to be not the best way to measure
  $H_0$) with cosmological parameter-values taken from other methods, as a way
  of constraining $P(g\,|\,F)$.
\end{enumerate}


%
%
%
\section*{Acknowledgments}

  We would like to thank Liliya L. R. Williams for useful discussions and
  comments on the paper.

  We also thank the anonymous referee for the constructive suggestions to
  bring the paper to its final form.

  PD acknowledges support from the Swiss National Science Foundation.  SM
  acknowledges the funding from the European Research Council (ERC) under the
  EUs Horizon 2020 research and innovation programme (COSMICLENS; grant
  agreement no. 787886).

  This research is based on observations made with the NASA/ESA Hubble Space
  Telescope obtained from the Space Telescope Science Institute, which is
  operated by the Association of Universities for Research in Astronomy, Inc.,
  under NASA contract NAS 5–26555. These observations are associated with
  programs \#10886, \#10174, \#12210, \#10494.

\section*{Data availability}
  The data underlying this article are available at the STScI
  (\href{https://mast.stsci.edu/}{https://mast.stsci.edu/}; the unique
  identifiers are cited in the acknowledgements).  The derived data generated in
  this research will be shared on request to the corresponding author, or can be
  replicated using the open-source software available at:
  \faGithub\;\href{https://github.com/phdenzel/gleam}{https://github.com/phdenzel/gleam}.

%
\bibliographystyle{mnras}
\bibliography{refs}

%
%

\newcommand{\onecaption}[1]{Results for SDSS~#1. See \secref{match:results} for details.}

\def\pwidth{.49\textwidth}
\def\hheight{.5\textheight}
\def\theight{.32\textheight}
\def\qheight{.23\textheight}

\newcommand{\lensmosaic}[2]{%
  \begin{tabular}{ll}
    &\includegraphics[#2]{\home/imgs/mcmc/#1_data}\includegraphics[#2]{\home/imgs/mcmc/#1_synth}\\%
    &\includegraphics[#2]{\home/imgs/mcmc/#1_srcplane}\includegraphics[#2]{\home/imgs/mcmc/#1_resid}\\%
    &\includegraphics[#2]{\home/imgs/mcmc/#1_kappa}\includegraphics[#2]{\home/imgs/mcmc/#1_pixrad11}\\%
    &\includegraphics[#2]{\home/imgs/mcmc/#1_arriv}\includegraphics[#2]{\home/imgs/mcmc/#1_kprofile}%
  \end{tabular}
}

\newcommand{\addmosaic}[2]{%
  \includegraphics[#2]{imgs/mcmc/{{SDSSJ0029-0055_#1}}}\includegraphics[#2]{imgs/mcmc/{{SDSSJ0737+3216_#1}}}\\%
  \includegraphics[#2]{imgs/mcmc/{{SDSSJ0753+3416_#1}}}\includegraphics[#2]{imgs/mcmc/{{SDSSJ0956+5100_#1}}}\\%
  \includegraphics[#2]{imgs/mcmc/{{SDSSJ1051+4439_#1}}}\includegraphics[#2]{imgs/mcmc/{{SDSSJ1430+6104_#1}}}\\%
  \includegraphics[#2]{imgs/mcmc/{{SDSSJ1627-0053_#1}}}%
}

\newcommand{\sidebysideA}[3]{
  \includegraphics[#3]{imgs/mcmc/{{SDSSJ0029-0055_#1}}}\includegraphics[#3]{imgs/mcmc/{{SDSSJ0029-0055_#2}}}\\%
  \includegraphics[#3]{imgs/mcmc/{{SDSSJ0737+3216_#1}}}\includegraphics[#3]{imgs/mcmc/{{SDSSJ0737+3216_#2}}}\\%
  \includegraphics[#3]{imgs/mcmc/{{SDSSJ0753+3416_#1}}}\includegraphics[#3]{imgs/mcmc/{{SDSSJ0753+3416_#2}}}%
}

\newcommand{\sidebysideB}[3]{
  \includegraphics[#3]{imgs/mcmc/{{SDSSJ0956+5100_#1}}}\includegraphics[#3]{imgs/mcmc/{{SDSSJ0956+5100_#2}}}\\%
  \includegraphics[#3]{imgs/mcmc/{{SDSSJ1051+4439_#1}}}\includegraphics[#3]{imgs/mcmc/{{SDSSJ1051+4439_#2}}}\\%
  \includegraphics[#3]{imgs/mcmc/{{SDSSJ1430+6104_#1}}}\includegraphics[#3]{imgs/mcmc/{{SDSSJ1430+6104_#2}}}\\%
  \includegraphics[#3]{imgs/mcmc/{{SDSSJ1627-0053_#1}}}\includegraphics[#3]{imgs/mcmc/{{SDSSJ1627-0053_#2}}}%
}

\newcommand{\triplecolA}[4]{
  \includegraphics[#4]{imgs/mcmc/{{SDSSJ0029-0055_#1}}}\includegraphics[#4]{imgs/mcmc/{{SDSSJ0029-0055_#2}}}\includegraphics[#4]{imgs/mcmc/{{SDSSJ0029-0055_#3}}}\\%
  \includegraphics[#4]{imgs/mcmc/{{SDSSJ0737+3216_#1}}}\includegraphics[#4]{imgs/mcmc/{{SDSSJ0737+3216_#2}}}\includegraphics[#4]{imgs/mcmc/{{SDSSJ0737+3216_#3}}}\\%
  \includegraphics[#4]{imgs/mcmc/{{SDSSJ0753+3416_#1}}}\includegraphics[#4]{imgs/mcmc/{{SDSSJ0753+3416_#2}}}\includegraphics[#4]{imgs/mcmc/{{SDSSJ0753+3416_#3}}}%
}

\newcommand{\triplecolB}[4]{
  \includegraphics[#4]{imgs/mcmc/{{SDSSJ0956+5100_#1}}}\includegraphics[#4]{imgs/mcmc/{{SDSSJ0956+5100_#2}}}\includegraphics[#4]{imgs/mcmc/{{SDSSJ0956+5100_#3}}}\\%
  \includegraphics[#4]{imgs/mcmc/{{SDSSJ1051+4439_#1}}}\includegraphics[#4]{imgs/mcmc/{{SDSSJ1051+4439_#2}}}\includegraphics[#4]{imgs/mcmc/{{SDSSJ1051+4439_#3}}}\\%
  \includegraphics[#4]{imgs/mcmc/{{SDSSJ1430+6104_#1}}}\includegraphics[#4]{imgs/mcmc/{{SDSSJ1430+6104_#2}}}\includegraphics[#4]{imgs/mcmc/{{SDSSJ1430+6104_#3}}}\\%
  \includegraphics[#4]{imgs/mcmc/{{SDSSJ1627-0053_#1}}}\includegraphics[#4]{imgs/mcmc/{{SDSSJ1627-0053_#2}}}\includegraphics[#4]{imgs/mcmc/{{SDSSJ1627-0053_#3}}}%
}

\begin{figure*}
    \centering
    \lensmosaic{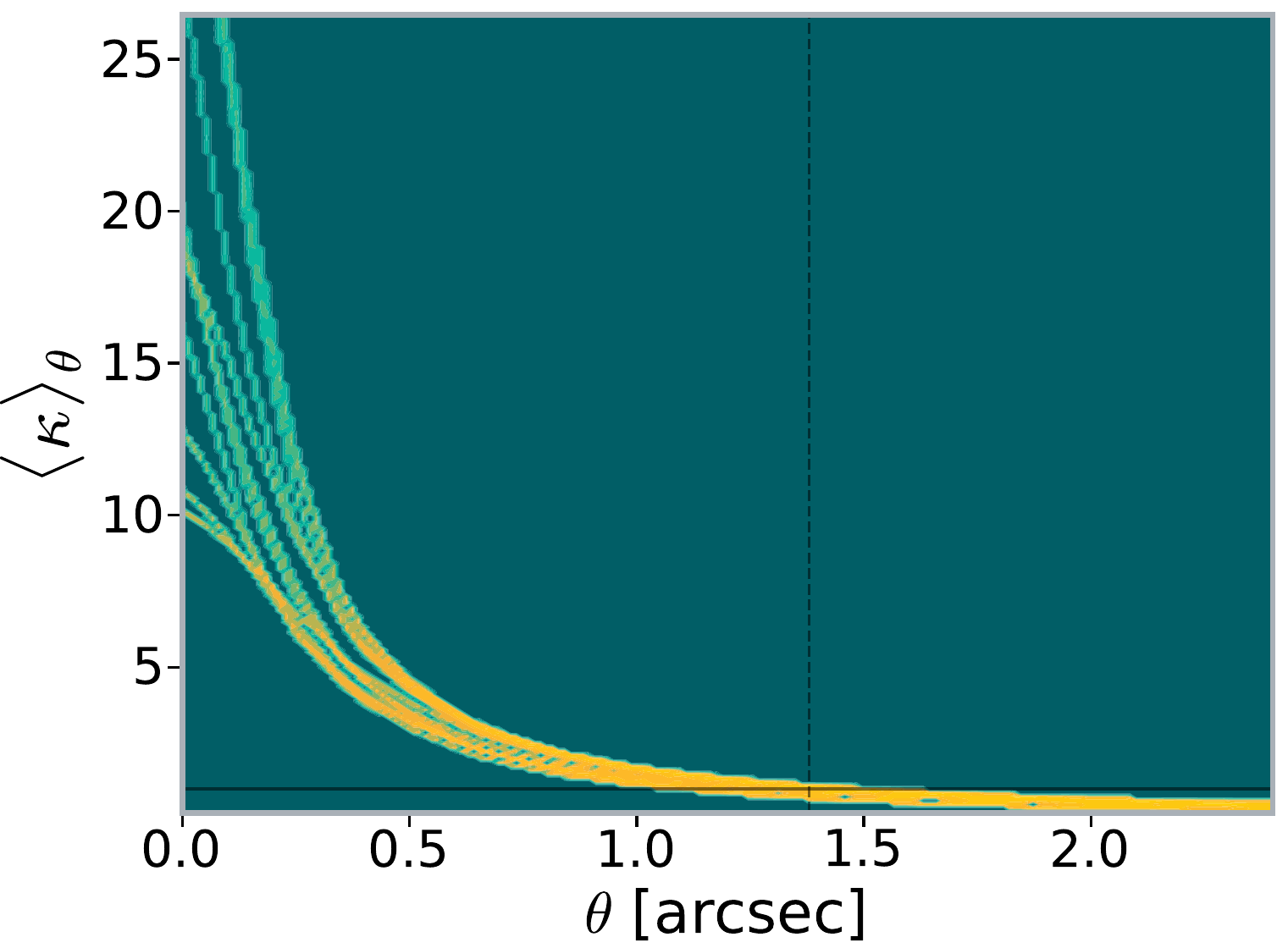}{height=\qheight}
    \caption{\onecaption{J0029-0055}}\figlbl{match:J0029}
\end{figure*}

\begin{figure*}
    \centering
    \lensmosaic{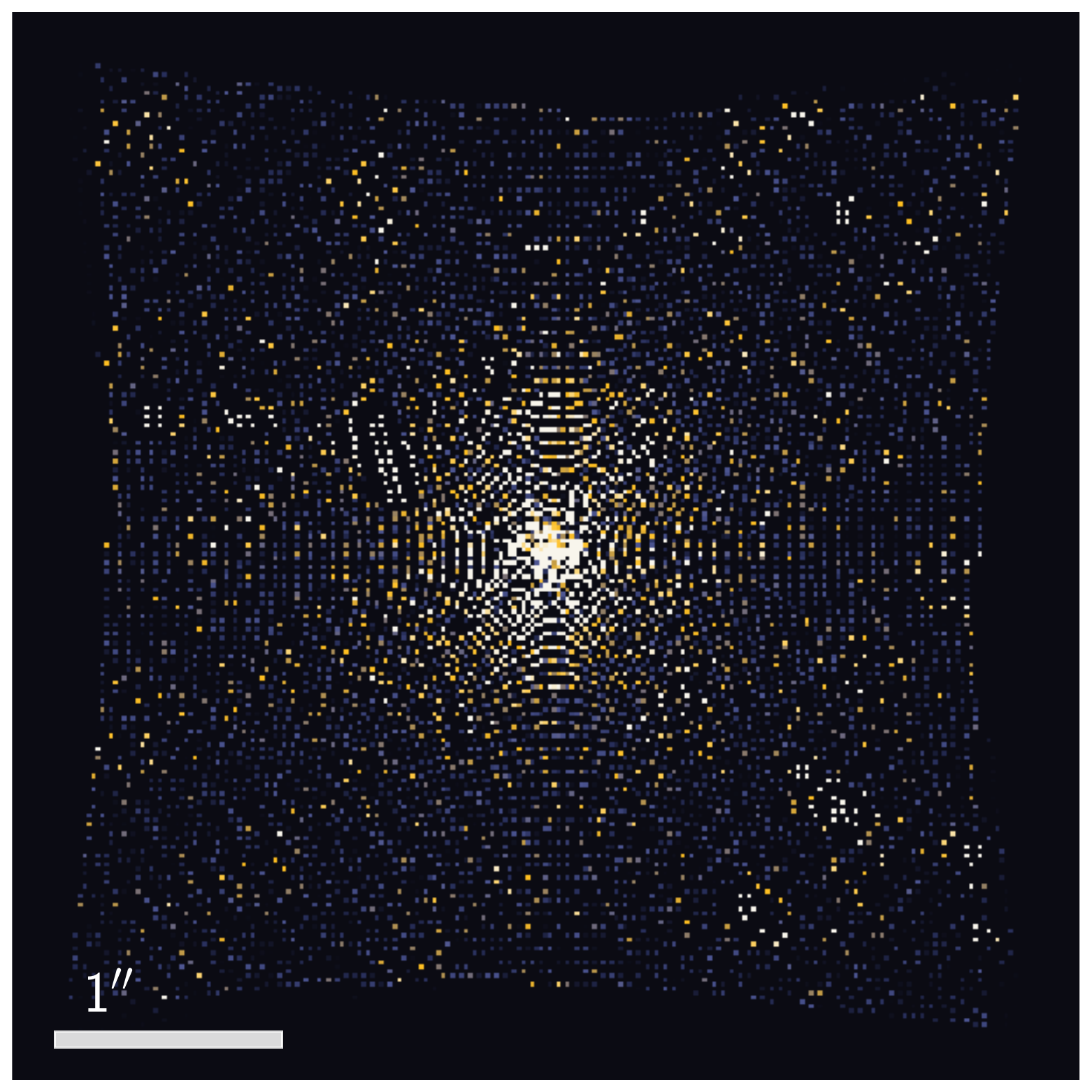}{height=\qheight}
    \caption{\onecaption{J0737+3216}}\figlbl{match:J0737}
\end{figure*}

\begin{figure*}
    \centering
    \lensmosaic{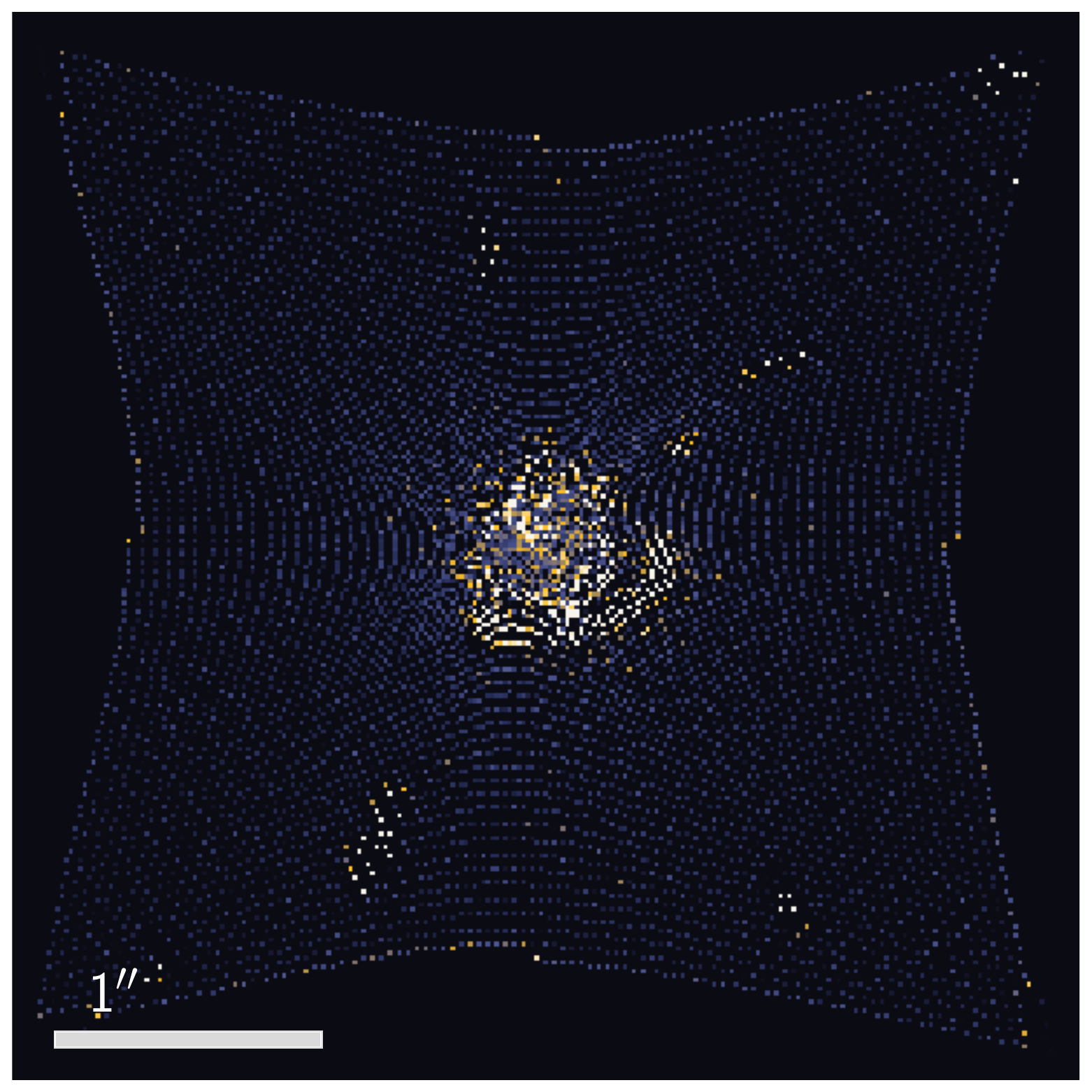}{height=\qheight}
    \caption{\onecaption{J0753+3416}}\figlbl{match:J0753}
\end{figure*}

\begin{figure*}
    \centering
    \lensmosaic{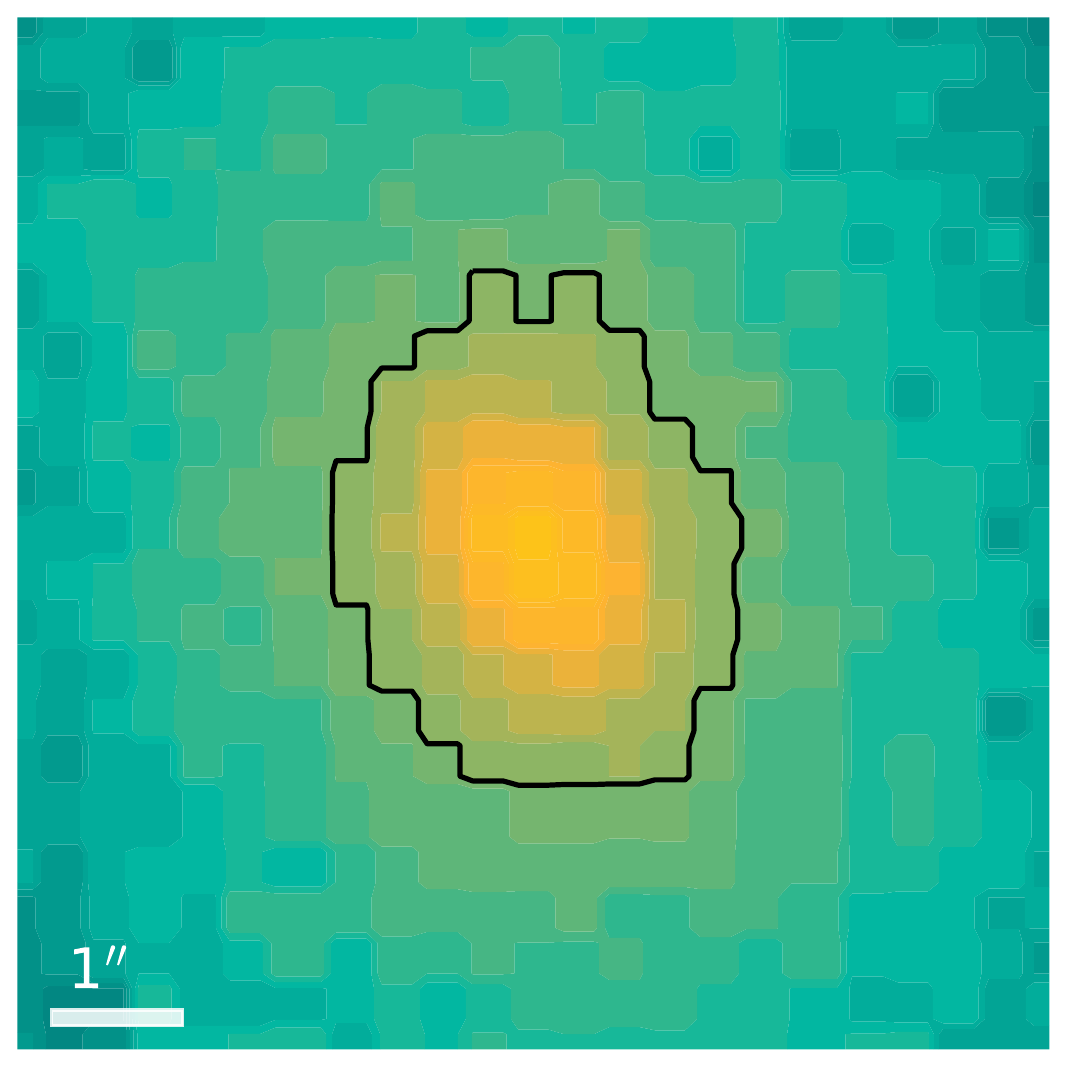}{height=\qheight}
    \caption{\onecaption{J0956+5100}}\figlbl{match:J0956}
\end{figure*}

\begin{figure*}
    \centering
    \lensmosaic{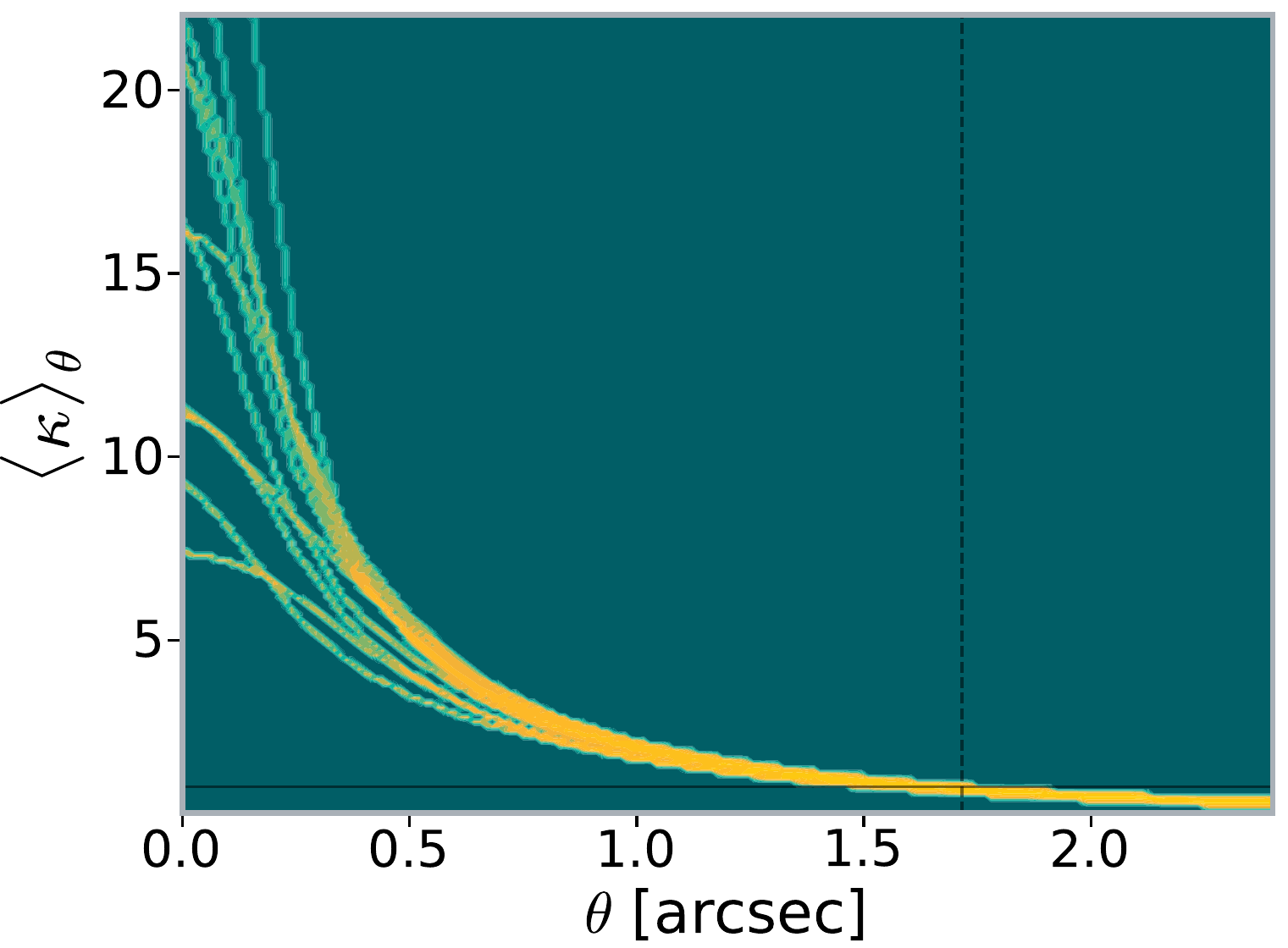}{height=\qheight}
    \caption{\onecaption{J1051+4439}}\figlbl{match:J1051}
\end{figure*}

\begin{figure*}
    \centering
    \lensmosaic{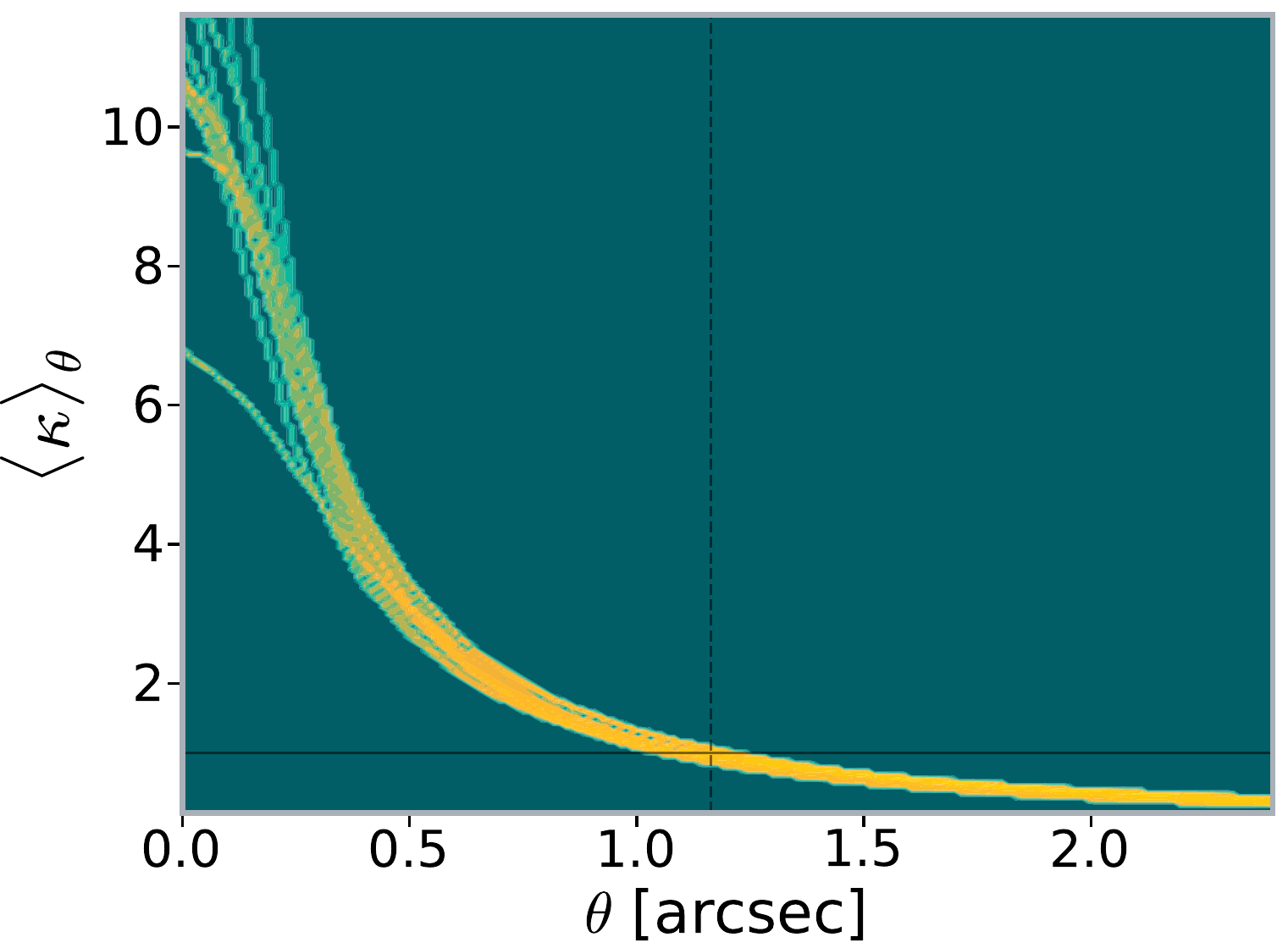}{height=\qheight}
    \caption{\onecaption{J1430+6104}}\figlbl{match:J1430}
\end{figure*}

\begin{figure*}
    \centering
    \lensmosaic{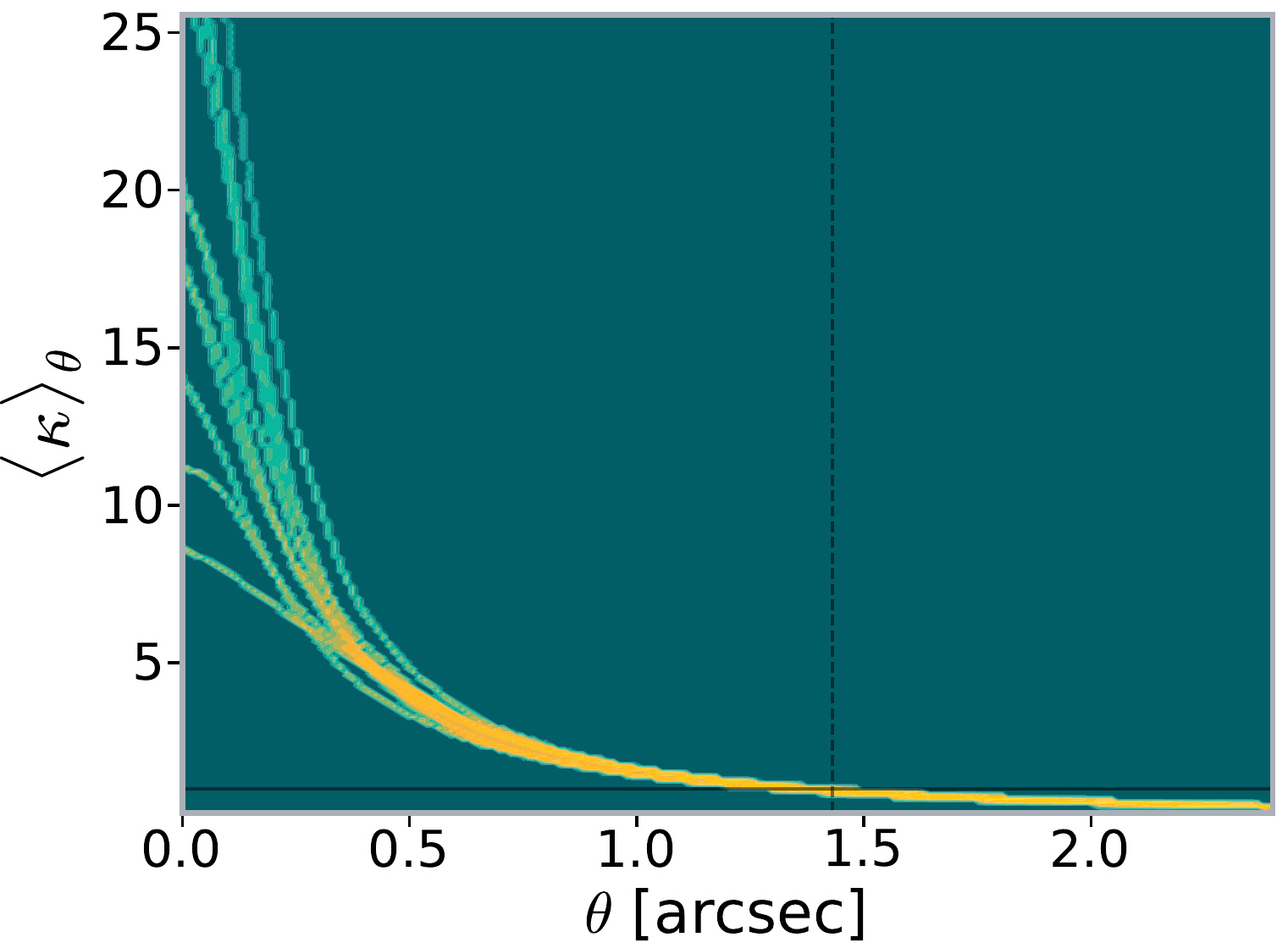}{height=\qheight}
    \caption{\onecaption{J1627-0053}}\figlbl{match:J1627}
\end{figure*}

\label{lastpage}
\clearpage

\end{document}